**Polar Nematic Phases with Enantiotropic Ferro- and Antiferroelectric Behavior**

*Mateusz Mrukiewicz, Michał Czerwiński*, Natalia Podoliak, Dalibor Repček, Paweł Perkowski, Richard. J. Mandle, Dorota Węgłowska*

M. Mrukiewicz, P. Perkowski
Institute of Applied Physics, Military University of Technology, Kaliskiego 2, 00-908 Warsaw, Poland

M.Czerwiński, D. Węgłowska
Institute of Chemistry, Military University of Technology, Kaliskiego 2, 00-908 Warsaw, Poland
E-mail: michal.czerwinski@wat.edu.pl

N. Podoliak, D. Repček
Institute of Physics, Academy of Science of the Czech Republic, Na Slovance 2, 182 00 Prague 8, Czech Republic

D. Repček
Faculty of Nuclear Sciences and Physical Engineering, Czech Technical University in Prague, Břehová 7, 110 00 Prague 1, Czech Republic

R. J. Mandle
School of Chemistry, University of Leeds, Leeds, UK, LS2 9JT
School of Physics and Astronomy, University of Leeds, Leeds, UK, LS2 9JT

The recent discovery of a new ferroelectric nematic ($N_F$) liquid crystalline phase became of utmost interest for the liquid crystal (LC) and the whole soft and condensed matter fields. Contrary to the previously known ferroelectric LC materials, whose ferroelectric characteristics were much weaker, new polar nematics exhibit properties comparable to solid ferroelectrics. This discovery brought about tremendous efforts to further explore compounds showing these phases, and fascinating physical properties have been reported. Herein, we present the first synthesized compounds with the enantiotropic ferro- ($N_F$) and antiferroelectric ($N_X$) nematic phases. The enantiotropic nature and an unprecedentedly broad temperature range of $N_F$ and $N_X$ phases are confirmed by various experimental techniques: polarized-light optical microscopy (POM) observations, different scanning calorimetry (DSC), dielectric spectroscopy, second harmonic generation (SHG), and molecular modeling.

The presented achievements in designing achiral compounds that exhibit enantiotropic polar nematic phases with ferro- and antiferroelectric properties significantly contribute to the development of multicomponent mixtures with a broad temperature range of $N_F$ and $N_X$ phases down to room temperature. Furthermore, this accomplishment considerably enhances the general understanding of the structural correlations that promote polar nematic liquid crystal phases with high thermodynamic stability. Finally, this work may benefit various applications in photonic devices.

**1. Introduction**

Incorporating nanotechnologies with electronics became widespread due to the use of components based on ferroelectric materials.[1,2] The technological need for enhanced properties in these materials has led to comprehensive scientific research in this field.[3–7] But nowadays, soft ferroelectrics can bring additional advantages by combining ferroelectric properties with fluidity.

Liquid crystals (LCs) belong to a field of soft matter, which uniquely combines the properties of both liquids and crystals. LC materials comprise organic molecules of anisotropic shape, which are packed with different degrees of orientational and positional order, forming various types of LC phases. These phases appear between the isotropic liquid (Iso) and crystalline (Cr) phases. Immediately below the isotropic phase, the less ordered nematic (N) LC phase could emerge. The average direction of molecules in nematic is characterized by a specific prevailing molecular orientation called director, *n*. Nematic shows the absence of the positional order. NLCs exhibit intriguing properties, including optical and dielectric anisotropy. These are crucial for light manipulation applications as they allow molecular orientation change in response to an applied electric field.

Below the N phase, more ordered lamellar LC phases, which are characterized by a certain degree of positional order in addition to orientational ones, could be found. These phases are called smectics, and they are distinguished by their molecular orientation within the layers. The less ordered smectic phase is smectic A (SmA), in which molecules are perpendicular to the layers. The more ordered smectic phase is smectic C (SmC), with molecules tilted relative to the layer normal. Besides, more ordered smectic phases could be found with additional positional molecular order within the layers but no correlation between the adjacent layers.

Initially, ferroelectric properties in LC materials, consisting of chiral molecules (SmC*), were discovered in the SmC phase.[8] Later, the antiferroelectric $SmC_A$* phase was also confirmed.[9–11] Since then, renewed interest in LC research arose as ferroelectric LCs promised new applicational potential. Ferro- and antiferroelectric behavior in LCs was initially assumed to be connected with molecular chirality, which cancels mirror symmetry. However, the situation changed with the discovery of polar phases and ferroelectricity in bent-shaped LCs, where the constituent molecules were not chiral.[12] Nonetheless, the scale of ferroelectricity in these materials was still far from the characteristics of solid ferroelectrics.

The existence of ferroelectric fluid with a large longitudinal dipole moment was predicted more than a century ago by Born.[13] However, not earlier than in 2017, the first such materials, with ferroelectric nematic phase ($N_F$), were synthesized simultaneously by two research groups; the materials were called RM734[14–16] and DIO.[17] Moreover, another polar nematic phase, termed here $N_X$, was also confirmed in the latter compound between the $N_F$ and N phases. Antiferroelectric features of the $N_X$ phase were observed,[18] and some researchers pointed to the smectic nature of this phase, referring to it as $SmZ_A$.[19]

The uniqueness of polar nematic phases is associated with their extraordinary properties. The $N_F$ phase shows enormous dielectric permittivity,[16,20–24] significant spontaneous polarization,[25] elevated non-linear optical activity,[26–31] and electro-optical response with an exceptionally low or near-zero threshold voltage.[20,32–37] Additionally, both $N_F$ and $N_X$ phases are characterized by notable fluidity. These exceptional properties hold potential for the development of unconventional soft matter applications.[26] Presented discoveries have become a stimulus for searching for new compounds and mixtures,[20,22,23,26,33,38–47] with the broadest possible range of the $N_F$ and/or $N_X$ phases, primarily exhibiting enantiotropic phase transitions. Nevertheless, until the previous year, all synthesized materials demonstrated the $N_F$ phase characterized by a monotropic nature, suggesting its potential thermodynamic metastability. Nishikawa et al.[48] explained this phenomenon as a consequence of the facile flipping of molecules from a synparallel arrangement, destabilizing the $N_F$ phase. This interpretation was derived from the analysis of the $N_F$ phase using single-crystal X-ray diffraction.[22,23,49]

However, our research group recently documented the first example of the compound, namely 3JK, exhibiting both $N_F$ and $N_X$ phases with an enantiotropic nature.[50] Despite this example, three fundamental classes of compounds have been identified, showing the $N_F$ and/or $N_X$ phases, albeit with monotropic transitions. Their primary representatives are presented in **Figure 1A**.

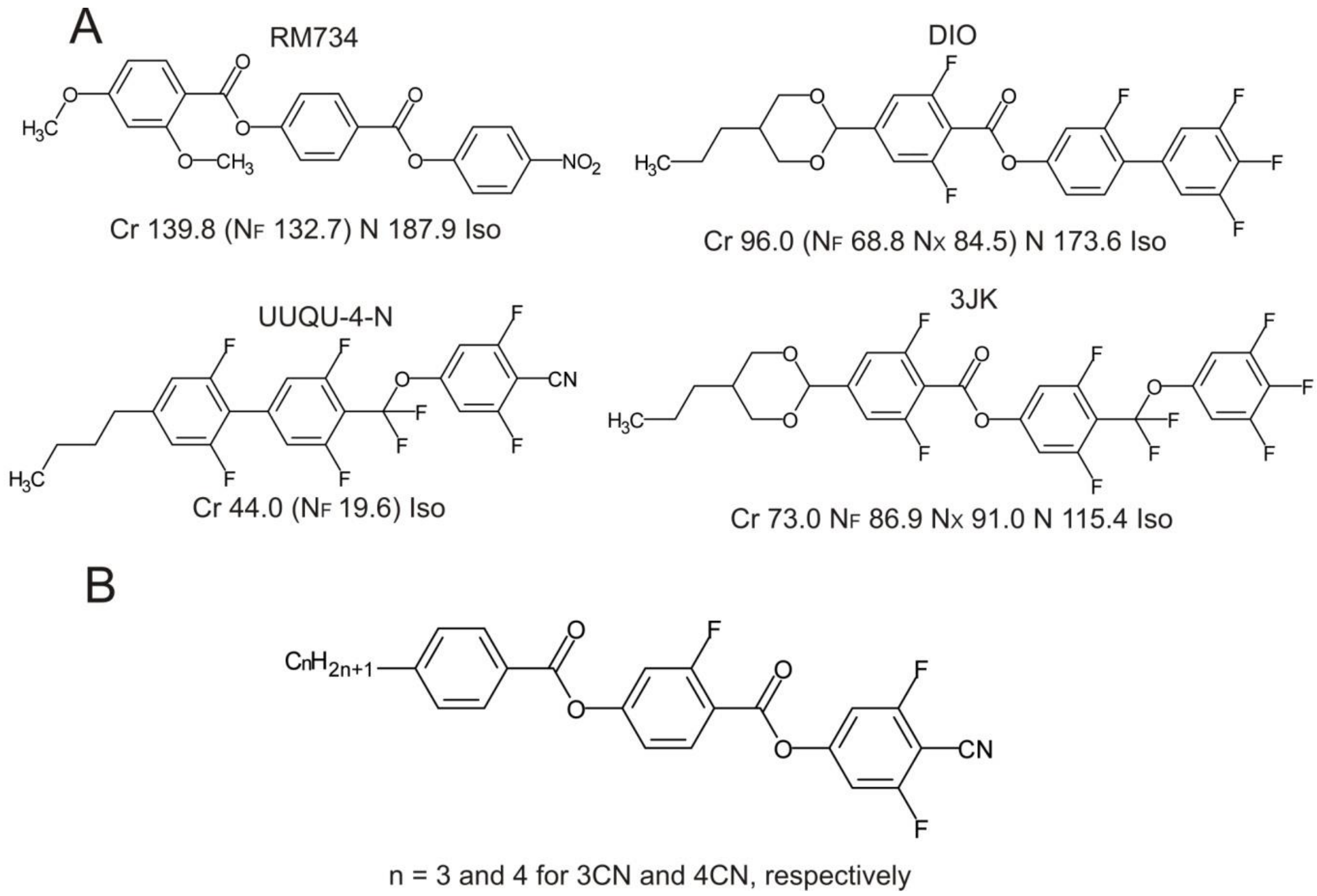

**Figure 1.** Chemical structure. A) Of compounds RM734,[14] DIO,[17] UQU-4-N,[21] and 3JK[50] with phase transition temperatures. B) Of compounds 3CN and 4CN.

Compound RM734 shows phase sequence Iso-N-$N_F$-Cr and consists of three benzene rings and two ester groups in the rigid core, two methoxy groups as substituents, and a terminated nitro group. Compound DIO, the first reported compound with both $SmZ_A$ and $N_F$ monotropic phases, exhibits three benzene rings, one dioxane ring, and one ester group in the rigid core as well as an alkyl and a fluorine atom as terminating elements and several fluorine atoms as lateral substituents. The direct Iso-$N_F$ phase transition is observed in the case of compound UUQU-4-N with three partially fluoro-substituted benzene rings and one fluorometoxy group in the rigid core, and alkyl and cyano groups as terminal elements. On the other hand, the 3JK compound manifests enantiotropic $N_F$ and $N_X$ phases, exhibiting chemical building units that can be classified into two types. The first type comprises an alkyl terminated group, 1,3-dioxane, and an aromatic ring substituted by two fluorine atoms and one ester group in an oriented fashion (DIO analog). The second type features a fluorocarbon ether group and fluorine atoms laterally substituted in two aromatic rings (UUQU-4-N analog), albeit with a fluorine atom instead of a cyano group as the terminating element.

Based on the presented structural analysis and other works regarding the search for the correlation between the mesomorphism and the structure in the compounds similar to those depicted in Figure 1A,[15,39,41,42] we examined our resources of liquid crystalline compounds with some structural similarity to the mentioned above. In works published in 2009,[51,52] we presented the synthesis and properties of fluoro-substituted three-ring phenyl esters with the cyano-terminated group. The structure of the studied compounds contains chemical elements of currently known compounds with polar nematic phases, such as: terminal cyano group, lateral fluorine atoms, and ester group linked benzene rings in the rigid core. At the time of the study, we identified several high-order phases below the N phase, classified as smectic phases, for two of them (here termed 3CN and 4CN). Since these studies and compounds were dedicated to the preparation of the nematic mixture for the dual frequency application[53,54] and pretransitional effect,[55] we did not conduct a deep analysis of such high-order phases. It is noteworthy that compounds 3CN and 4CN exhibit: three benzene rings as all of the representatives depicted in Figure 1; an alkyl terminated group and lateral substituted fluorine atoms as DIO, UUQU-4-N, and 3JK; a cyano-terminated group as UUQU-4-N; and two oriented in one direction ester groups between the benzene rings as RM734.

Herein, we present a deep investigation conducted to confirm the existence of polar nematic phases in the 3CN and 4CN, incorporating these materials into the first synthesized ferroelectric nematics. In this work, we report on mesomorphic, dielectric, electro-optical, and second harmonic generation (SHG) properties of 3CN and 4CN compounds belonging to the series of 4-cyano-3,5-difluorophenyl 4-(4-alkylbenzoyloxy)-2-fluorobenzoates, with the chemical structure presented in Figure 1B. Additionally, we study the mesomorphic properties of a mixture comprising 20.0 wt% of compounds 3CN and 80.0 wt% of DIO (3CN-DIO). The results unambiguously indicate that both compounds show enantiotropic phase transitions to the $N_X$ phase, and one of the compounds also shows a transition to the $N_F$ phase during heating. An unprecedentedly broad temperature range distinguishes both enantiotropic polar phases. Additionally, we confirmed the ferroelectric and antiferroelectric features of $N_F$ and $N_X$ phases, respectively. Considering 2009 as the documented date of the synthesis of 3CN and 4CN,[51] they can be regarded as the first compounds exhibiting ferro- and antiferroelectric phases upon melting. This work demonstrates that it is possible to achieve thermodynamically stable $N_F$ and $N_X$ phases, with a temperature range spanning several tens of Celsius degrees, by appropriately designing the molecular structure of the NLC compound.

## 2. Results and Discussion

First, the phase identification and the transition temperatures of 3CN, 4CN and mixture 3CN-DIO were determined using standard techniques such as polarized-light optical microscopy (POM) and differential scanning calorimetry (DSC). The temperatures, the enthalpy changes of the phase transitions, and the textures of liquid crystal phases of compounds 3CN, 4CN, and mixture 3CN-DIO are presented in **Table 1** and **Figure 2A**, respectively. Detailed DSC thermograms with integrated peaks are given in **Figure S1** (Supporting Information). The motivation for exploring the mixture of DIO with only 3CN lies in the enantiotropic nature observed for the $N_F$ and $N_X$ phases of 3CN.

**Table 1.** The phase transition temperatures [°C] (onset point), and corresponding enthalpy changes [kJ mol$^{-1}$], in italic font, as well as the temperature range of the ferroelectric nematic phase, $N_F$, and antiferroelectric nematic phase, $N_X$, [°C] of the members of the homologous series nCN, DIO compound[17] and 3CN-DIO mixture from DSC measurements determined during heating (upper rows) and cooling (down rows); values given in brackets were determined for monotropic phase.

| **Acronym** | **Cr** | **[°C]** ***[kJmol$^{-1}$]*** | **$N_F$** | **[°C]** ***[kJmol$^{-1}$]*** | **$N_X$** | **[°C]** ***[kJmol$^{-1}$]*** | **N** | **[°C]** ***[kJmol$^{-1}$]*** | **Iso** | **$N_F$ range [°C]** | **$N_X$ range [°C]** |
|---|---|---|---|---|---|---|---|---|---|---|---|
| 3CN | • | 109.6 | • | 156.2 | • | 168.9 | • | 189.5 | • | 46.6 | 12.7 |
| | | *25.28* | | *0.59* | | *0.03* | | *0.76* | | | |
| | | 39.3 | | 157.7 | | 173.9[a] | | 194.7 | | 118.4 | 16.2 |
| | | *-9.94* | | *-0.56* | | *-0.05* | | *-0.66* | | | |
| 4CN | • | 115.0 | (•) | - | • | 137.1 | • | 175.7 | • | - | 22.1 |
| | | *35.76* | | | | *0.02* | | *0.56* | | | |
| | | 80.2 | | (102.8) | | 137.3 | | 176.9 | | 22.6 | 34.5 |
| | | *-23.89* | | *(-0.39)* | | *-0.02* | | *-0.67* | | | |
| DIO[17] | • | 96.0 | (•) | - | (•) | - | • | 173.6 | • | - | - |
| | | *24* | | | | | | *0.40* | | | |
| | | 34.0 | | (68.8) | | (84.5) | | 173.6 | | 34.8 | 15.7 |
| | | *-14* | | *(-0.2)* | | *(-0.003)* | | *-0.40* | | | |
| 3CN-DIO | • | 86.4[b] | • | 99.3 | (•) | - | • | 176.5 | • | 12.9 | - |
| | | *12.46[b]* | | *0.25* | | | | *0.39* | | | |
| | | - | | 98.1 | | (112.6) | | 175.8 | | >78.1 | 14.5 |
| | | | | *-0.25* | | *(-0.01)* | | *-0.39* | | | |

[a]The value of the peak top due to a broad temperature range of the N-$N_X$ phase transition (see the inset in Figure 4b); [b]The values for the phase transition between second crystalline form ($Cr_2$) and ferroelectric nematic phase ($N_F$)

The melting and clearing temperatures of the studied compounds are below 120.0 °C and 190.0 °C, respectively. The elongation of the alkyl terminal chain in homologs of the nCN series increases the former temperature and decreases the latter. Additionally, it leads to a significant increase in the enthalpy change at the melting point, as shown in Table 1. During the first heating from the crystalline phase, the $N_F$ and $N_X$ phases appear in 3CN within 46.6 degrees and 12.7 degrees, respectively. The phase sequence during cooling remains the same, but the temperature range of the $N_F$ phase widens considerably. In the case of 4CN, the $N_F$ phase is observed only during cooling, making it a monotropic phase. However, the $N_X$ phase appears either during heating and cooling in 4CN, with temperature ranges of 22.1 degrees and 34.5 degrees, respectively.

The characteristic peaks observed in the DSC curves (Figure 2B-E; Figure S1, Supporting Information), as well as the values of the enthalpy changes presented in Table 1 for the phase transitions, along with the textures of the phases, provide confirming evidence for the mesomorphic properties described above in 3CN and 4CN. The enthalpy changes corresponding to the phase transitions of the studied compounds are two times larger, or even ten times, for the $N_X$ - $N_F$ phase transition compared to those of DIO. These differences can be attributed to the distinct chemical structure of the compared compounds and the variations in the temperatures at which the phase transitions occur. The entropy change is a characteristic parameter indicating the alteration in molecular order during a phase transition, expressed as the enthalpy per temperature. Notably, the $N_X$ - $N_F$ phase transition of 3CN and 4CN occurs at significantly higher temperatures than that of DIO, leading to higher enthalpy change values for this specific phase transition in the studied compounds.

The mesomorphic properties of mixture 3CN-DIO further confirm the phase sequence in the compound 3CN. The addition of 20.0 wt% of the compound 3CN to DIO induces the enantiotropic $N_F$ phase, and there is a significant increase in the upper temperature at which the $N_F$ phase and the $N_X$ phase appear during cooling. This evidence is supported by the data in Table 1 and Figure 2C and E. Furthermore, the temperature range of the $N_X$ phase upon cooling in mixture 3CN-DIO is between this range in compound 3CN and DIO. POM images revealed marble textures of the nematic phase of all studied materials (Figure 2B-a, e, i). During cooling, hazier and more wrinkled texture with more defect lines occurs at $N_X$ phase of 3CN (Figure 2A-b), 4CN (Figure 2A-f), and 3CN-DIO (Figure 2A-j). At the phase transition from the $N_X$ to $N_F$ phase, a drastic change with the spontaneous flow in the texture is observed. Finally, the $N_F$ phase is characterized by a typical multidomain texture with two opposite colors for all materials (Figure 2A-d, h, j). The video with the phase transitions of 3CN is given in Supporting Information (**Video S1**).

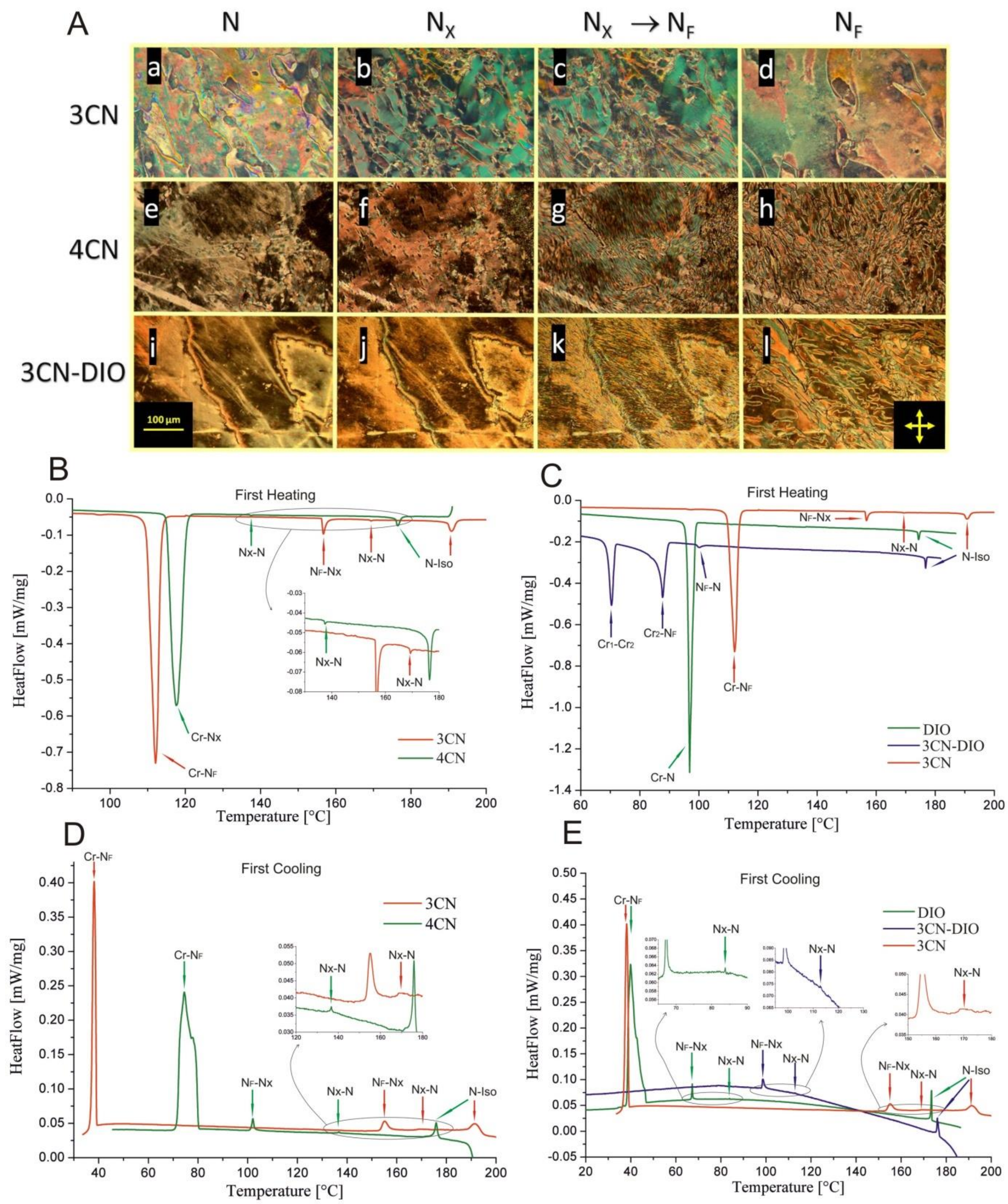

**Figure 2.** Mespomorphic properties of 3CN and 4CN compounds. A) POM images, obtained for material between not-rubbed glass plates on cooling from the isotropic phase, of 3CN (a-d), 4CN (e-h), and mixture DIO+0.2_3CN (i-l) under crossed polarizers: the N phase at 175.0 °C (a), 145.0 °C (e) and 125.0 °C (i); the $N_X$ phase at 165.0 °C (b), 120.0 °C (f) and 105.0 °C (j); the $N_X$ -$N_F$ phase transition at 157.5 °C (c), 102.5 °C (g) and 98 °C (k) – characteristic lines with spontaneous flow appear; the $N_F$ phase at 140.0 °C (d), 90.0°C (h) and 85.0 °C (l). B, C) Comparative DSC curves of individual compounds and mixture during the first heating. D, E) Comparative DSC curves of individual compounds and mixture during the first cooling.

The dielectric and electro-optical measurements in planarly aligned materials were performed further to confirm the ferroelectric behavior of 3CN and 4CN and fully characterize the intermediate antiferroelectric phase between the ferroelectric and paraelectric nematic. The phase transition temperatures in a confined volume differ from those obtained in bulk, which is a normal phenomenon for LC compounds. Upon cooling of the isotropic phase, the planarly aligned nematic phase appears both in 3CN and 4CN (**Figure 3A** and **G**). The transition to the intermediate phase is accompanied by a spontaneous flow with a small change of birefringence and the formation of poorly visible irregular stripes along the rubbing direction (Figure 3B and H). The $N_F$ phase of the investigated compounds is characterized by the typical lens-shaped domain texture.[25,32] The domains of opposite polarity are separated by visible disclinations lines (Figure 3C and I). The ferroelectric domains are switchable by the AC and DC fields (**Video S2** and **S3**, Supporting Information), as was presented earlier for other compounds with the monotropic $N_F$ phase.[25,32,35] The optical studies confirm that the 3CN compound possesses the ferroelectric nematic phase in the heating cycle from the crystal phase (Figure 3D and E). The $N_F$ phase created during the heating cycle shows many more defects than the one created in the cooling cycle. The $N_X$ phase is visible during heating under the POM in 3CN and 4CN. However, the texture of the $N_X$ phase is different depending on the phase sequence in the heating cycle (Figure 3F and K). All materials exhibit high-temperature paraelectric nematic phases (Figure 3L).

In the $N_X$ phase, the regions with different optical properties are observed (Figure 3M-O). Commencing from the cross-polarized position, in which the rubbing direction is at an angle of 45 degrees to the polarizer axis, we started moving the sample. Rotating the sample clockwise and anticlockwise with the cell at the same angle (20 degrees) causes the appearance of light and dark domains formed in periodic stripes. Examples of domains are marked in blue and green circles (Figure 3N and O). In the experiment, the previously dark domains in one position become light in the second position and vice versa. The effect arises from domains with different orientations of optical axes, which may indicate the antiferroelectric nature of the $N_X$ phase.

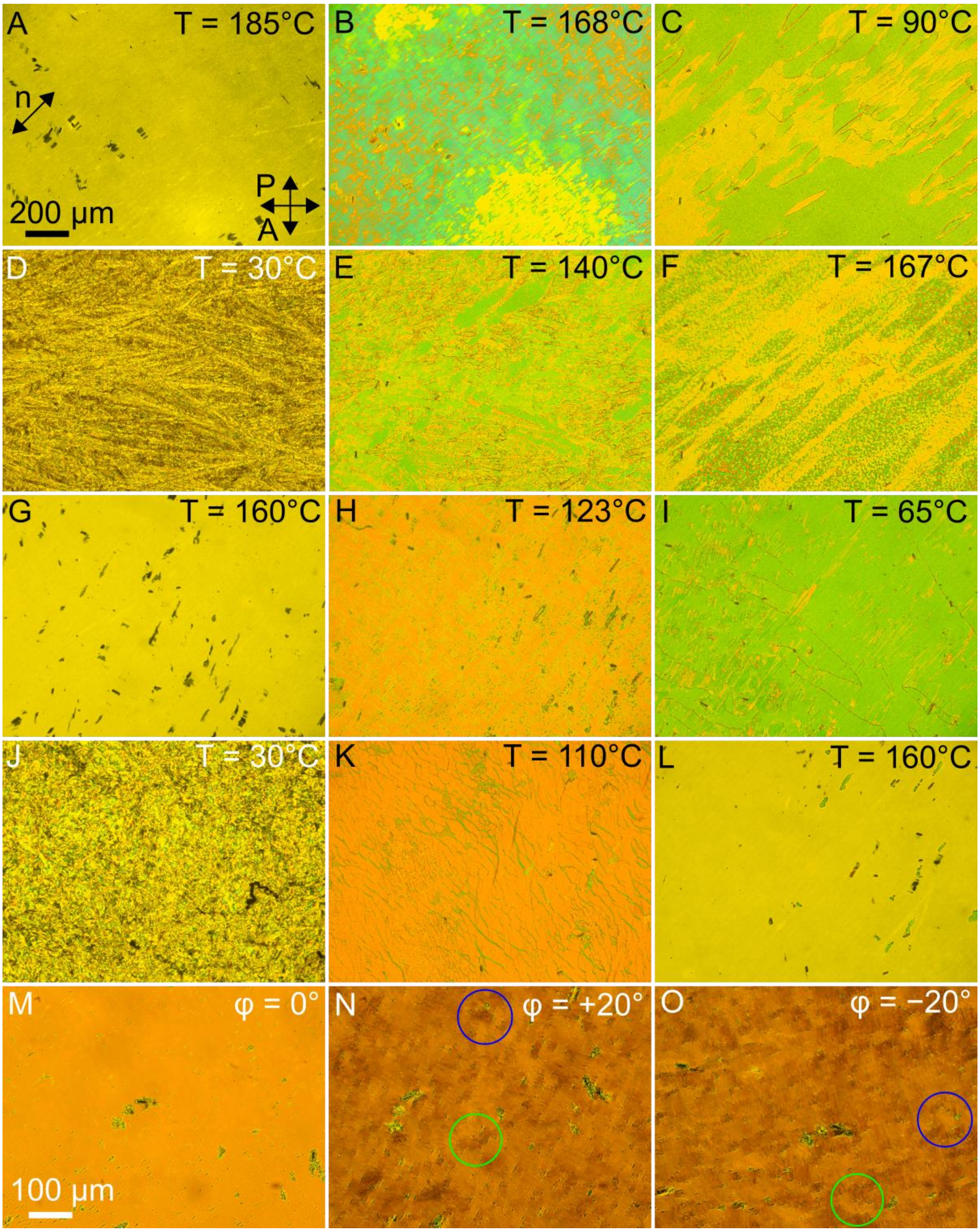

**Figure 3.** POM images of planarly aligned materials under crossed polarizers. A-F) For 3CN in the cooling cycle (A-C) and the heating cycle (D-F). Textures were obtained in: the N phase (A), the $N_X$ phase (B), the $N_F$ phase (C), the Cr phase (D), the $N_F$ phase (E), the $N_X$ phase (F). G-L) For 4CN in the cooling cycle (G-I) and the heating cycle (J-L). Textures were obtained in: the N phase (G), the $N_X$ phase (H), the $N_F$ phase (I), the Cr phase (J), the $N_X$ phase (K), the N phase (L). M-N) Textures of $N_X$ observed at 115°C before rotating (M), by rotating the sample by 20 degrees clockwise (N) and anticlockwise (O).

To confirm the enantiotropic character of polar nematic phases and their ferroelectricity or antiferroelectricity, the dielectric spectroscopy measurements over a wide range of frequencies (100 Hz – 10 MHz) and temperatures (30°C – 200°C) were performed. **Figure 4A-D** presents the temperature dependence of the real part of electric permittivity $\varepsilon'$ of 3CN and 4CN compounds in the cooling and heating cycles. In the cooling of 3CN, we can distinguish three mesophases, starting from high temperatures: paraelectric nematic phase N, the antiferroelectric nematic phase $N_X$, and the ferroelectric nematic phase $N_F$. The $N_F$ phase is characterized with much higher electrical permittivity (650÷850) at low frequencies than other phases. However, the observed values are lower than for RM 734[25,56–58] but similar to the results obtained for DIO.[56] Unexpected electric response behavior is noticed at the Iso-N phase transition where, despite the planar alignment, the electric permittivity increases ($\varepsilon'$ ~ 450 at 1 kHz in N). Subsequently, the electric permittivity suddenly begins to drop rapidly, which is associated with the appearance of the $N_X$ phase (N-$N_X$ transition). At the $N_X$-$N_F$ phase transition $\varepsilon'$ starts to increase again, which is related to the appearance of the polar ferroelectric domains. The dielectric spectroscopy measurements in the heating cycle confirm the enantiotropic behavior of the $N_F$ and $N_X$ phases in 3CN (Figure 4C). The ferroelectric nematic phase is formed at 113°C from the crystal (Cr) phase, where the value of $\varepsilon'$ is around 3. The Cr phase in the investigated materials was created and relaxed for 24 hours at room temperature after the cooling cycle. The created $N_F$ phase in the heating process exhibits different frequency dispersion characteristics than in cooling due to the smaller size of the ferroelectric domains and the disorder of the molecular orientation. Again, the $N_X$ phase is characterized with a lower value of $\varepsilon'$ than that can be observed in $N_F$, but significantly higher than in the N phase.

In contrast to 3CN, the 4CN compound does not exhibit the enantiotropic nature of the $N_F$ phase. However, the antiferroelectric phase, $N_X$, is observed in both cycles of experiments. In the cooling cycle of 4CN, the transition to the $N_X$ phase is also visible as the decrease in electric permittivity at the N-$N_X$ phase transition (Figure 4B). In this case, the antiferroelectric phase is much broader and occurs in the lower temperatures range, from 137°C to 89°C, compared to that in 3CN. Close to the $N_X$-$N_F$ phase transition, the frequency dispersion of electric permittivity is small, indicating the occurrence of ferroelectric relaxations in a similar frequency range. A similar phenomenon is observed in the heating cycle slightly above 100°C (Figure 4D), which proves the existence of the same phase. It is worth emphasizing that Figures 4B and 4D do not show any significant difference in the dielectric behavior of the nematic phase created in the cooling and heating cycles, which also proves the existence of the enantiotropic $N_X$ phase.

The decrease in electric permittivity in 3CN and 4CN, within the temperature ranges of the $N_X$ phase, is due to the creation of the antiferroelectric phase. The observed transition from the paraelectric phase to the antiferroelectric phase causes reducing the dielectric response due to the creation of the domains with opposite dipole moments. A similar pronounced decrease in $\varepsilon'$ was also reported in chiral liquid crystals when a ferroelectric smectic SmC* transforms into an antiferroelectric smectic $SmC_A$* phase.[57,58] This effect is caused by the nearly antiparallel orientation of dipoles in adjacent smectic layers. Due to this fact, the dipolar contribution to an electric response of $SmC_A$* is almost canceled. In the case of paraelectric-antiferroelectric transition in the investigated compounds, electric permittivity declines from ~275 to ~125 at 1 kHz for 4CN, whereas in polar smectics at the ferro-antiferroelectric transition (SmC*-$SmC_A$*) $\varepsilon'$ changes from ~200 to ~10. The disappearance of the Goldstone mode when entering the $SmC_A$* phase is responsible for the pronounced decrease in dielectric constant in chiral smectics.[59,60] Here, the sudden changes in the dielectric constant between the phases can also be related to the existence or disappearance of the dielectric modes, which will be discussed later.

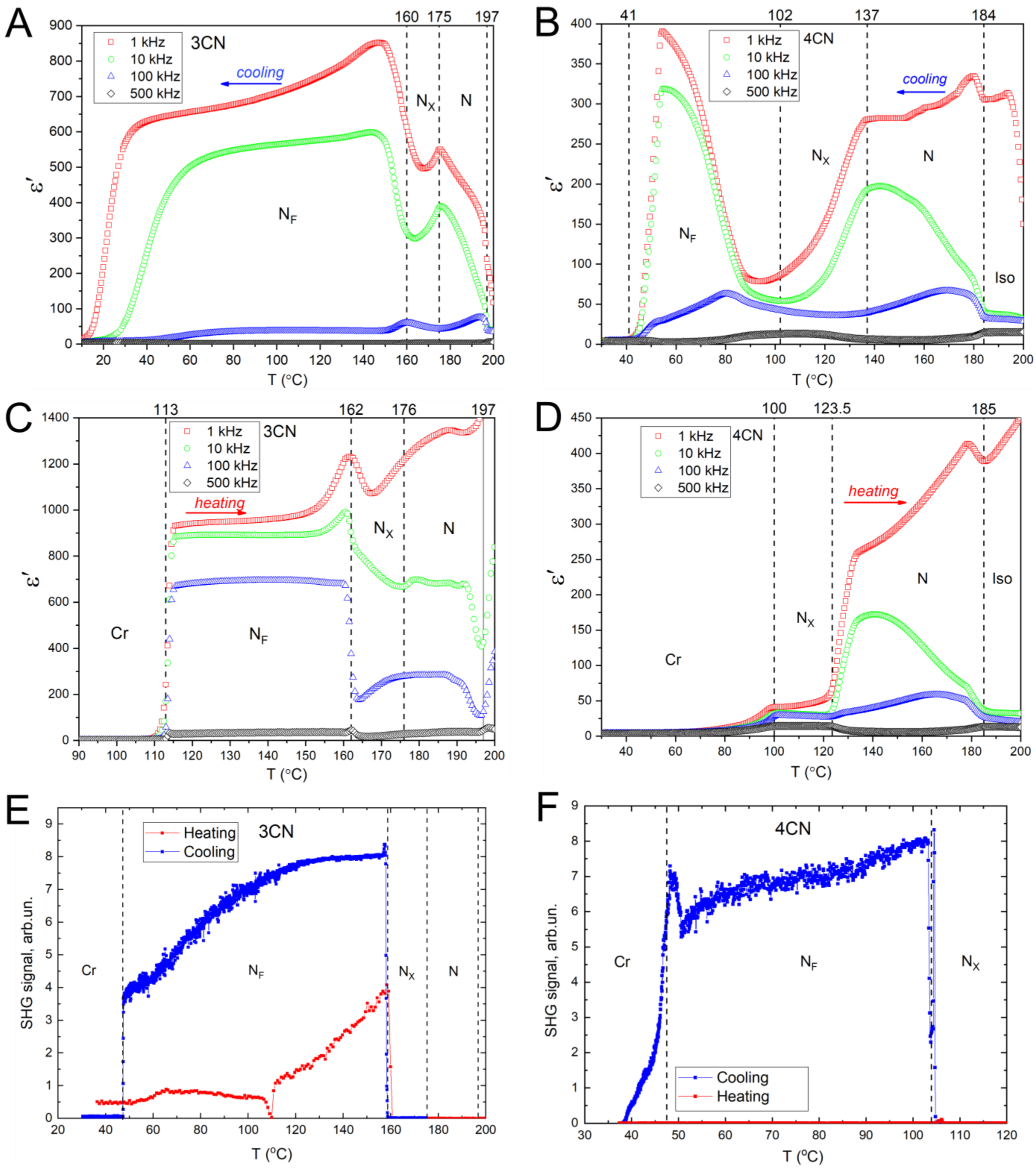

**Figure 4.** Dielectric and SHG features. The phases are designated, and the phase transition temperatures in the cooling cycle are marked with dashed lines. Real part of electric permittivity $\varepsilon'$ at frequencies of 1, 10, 100, 500 kHz versus temperature *T*. A, B) In the cooling cycle for 3CN (A) and 4CN (B). C, D) In the heating cycle for 3CN (C) and 4CN (D). E, F) Temperature dependence of the SHG signal for 3CN (E) and 4CN (F). The results are presented for the measurements performed on heating (red curve) and cooling (blue curve).

SHG measurements were performed with the aim of proving the phase sequence and the polarity of the $N_F$ phase. The experiments were conducted in a transmission mode according to the scheme described in the experimental section. The results are presented in Figure 4E and F both in heating and cooling cycles for 3CN (Figure 4E) and 4CN (Figure 4F). For **3CN**, the SHG signal shows non-zero values in the $N_F$ phase both on heating and cooling, proving the enantiotropic character of the ferroelectric phase. However, the values observed on heating are lower than the ones observed on cooling. This might be due to the remaining of the certain type of crystalline features. The temperature range, at which the SHG shows non-zero values, is narrower on heating as the transition from the crystalline to the $N_F$ phase occurs at a higher temperature. With the transition to the $N_X$ phase, the SHG signal abruptly falls down to zero, proving the absence of polarity in this phase. The SHG signal shows zero values within the Iso, N, and $N_X$ phases on cooling from the isotropic phase. Then, it rapidly grows at the $N_X$-$N_F$ phase transition to its maximum values, possessing a sharp "tooth" at the transition point. Within the $N_F$ phase, the signal slowly decreases. A step-like drop down to zero is observed at the crystalline transition.

For 4CN, the second harmonic is not generated during the measurements carried out on heating from the crystalline phase, and the signal remains zero. For the measurements conducted on cooling, the SHG signal shows zero values within the Iso, N, $N_X$ phases, and a sharp increase is observed at the transition to the $N_F$ phase. It is worth pointing out that for 4CN the "tooth" observed at the transition is even more pronounced than for 3CN, followed by a decrease in the SHG values and then by a rapid increase back up to almost similar values. Within the temperature range of the $N_F$ phase, a slight decrease in SHG values is observed with the temperature lowering. At the crystallization transition, another "tooth" appears. This behavior of the SHG signal at the phase transitions might be connected with the pretransitional instabilities of the system.

To further confirm the antiferroelectric behavior of the $N_X$ phase, we compared the dielectric spectroscopy results of 4CN with DIO. The intermediate phase of DIO, was confirmed as the antiferroelectric phase between the paraelectric and ferroelectric phases.[19] In comparison to 3CN, the 4CN compound has a broad and low-temperature $N_X$ phase, which permittivity dispersion is similar to the characteristics of DIO.[56] **Figures 5A** and **B** present the 3D plot of real $\varepsilon'$ and imaginary $\varepsilon''$ parts of the electric permittivity versus temperature and frequency. The dielectric spectroscopy measurements show that the existence of the $N_X$ phase[56] is related to the detection of two well-separated relaxations (high- and low-frequency dielectric modes) in the measured frequency domain. The formation of the antiferroelectric phase begins in the nematic phase with the appearance of the high-frequency mode at the vicinity of the N-$N_X$ phase transition. This process is visible in Figures 4B and 5A as the increase in electric permittivity in the N phase of 4CN. This phenomenon was also visible in DIO.[56] The transition to $N_F$ is characterized by a sharp change in the dielectric response due to the appearance of strong ferroelectric modes of relaxation frequencies around 10 kHz. In the case of 4CN and 3CN, the ionic mode at frequencies around 0.1 kHz is also detected in the dielectric spectrum. Because the ionic relaxation is not strictly connected to the ferroelectric or antiferroelectric behavior, it is not the subject of the research.

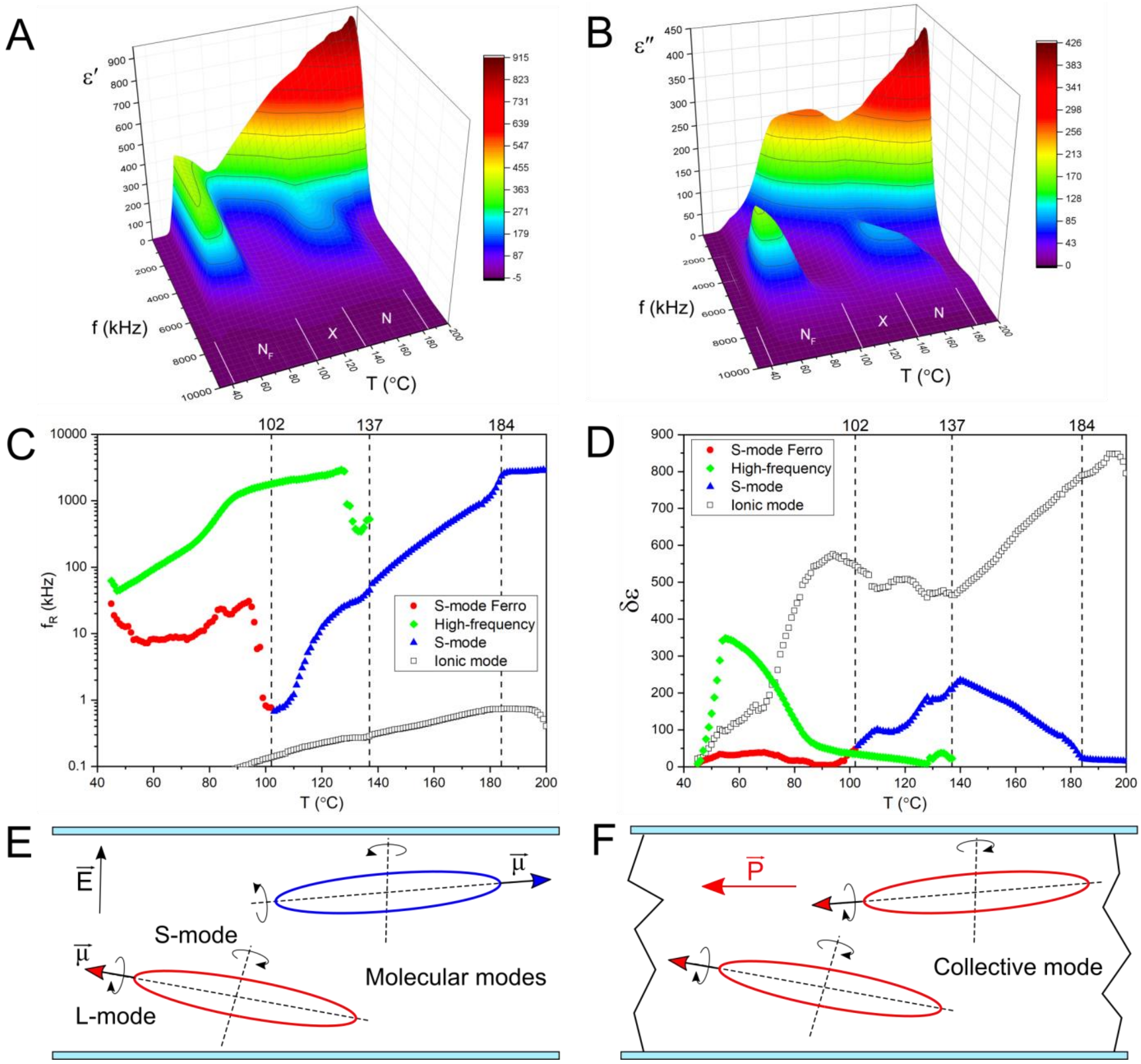

**Figure 5.** Deep dielectric analysis of the 4CN compound. A, B) Temperature and frequency dependence of real $\varepsilon'$ (A) and imaginary $\varepsilon''$ (B) part of electric permittivity measured across the isotropic phase, paraelectric nematic phase N, antiferroelectric phase $N_X$, ferroelectric nematic $N_F$ phase and crystal phase Cr. C, D) Temperature dependence of relaxation frequency $f_R$ (C) and dielectric strength $\delta\varepsilon$ (D) for detected dielectric modes. E, F) Relaxation mechanism of highly polar molecules in the N phase (E) and the $N_F$ phase(F).

The relaxation parameters of the dielectric modes observed in 4CN: the relaxation frequency $f_R$, and the dielectric strength $\delta\varepsilon$ were extracted from the Cole-Cole model[61] and plotted versus temperature (Figure 5C and D) for all visible phases. In the Iso and N phase, we observe only one dielectric mode - the relaxation around a short molecular axis, abbreviated as S-mode (Figure 5E). The rotation around the long molecular axis (L-mode), in the planarly aligned nematic phase typically remains undetectable by impedance analyzers due to the high-frequency range of detection, spanning hundreds of MHz to a few GHz. The S dielectric mode behaves as a typical molecular relaxation with $f_R$ decreasing and $\delta\varepsilon$ increasing with temperature. At the N-$N_X$ phase transition, a weak high-frequency relaxation appears. This phenomenon causes a significant decrease in $f_R$ and $\delta\varepsilon$ of the S-mode. As the temperature decreases, the dielectric strength of the high-frequency mode amplifies. The identical frequency and strength correlations were reported for $N_X$ in DIO.[56] The difference in relaxation parameters comes from different molecular structures of DIO and 4CN and the temperature ranges of the measurements. However, the similarities in the electric permittivity dispersion and molecular relaxation parameters are visible. In the ferroelectric phase, because of different molecular packing, the nature of the S-mode is different compared to the classical nematic phase (see S-mode Ferro, Figure 5C and D). The relaxation frequency and dielectric strength are relatively constant with the temperature. The high-frequency mode observed in the $N_X$ phase must be related to the domain structure of the $N_F$ phase because the mode is continued in the ferroelectric phase. This mode is probably a collective relaxation inside the ferroelectric domains. In the collective mode molecules with a dipole moment $\mu$ rotate in and out of phase in the ferroelectric neighboring domains with opposite polarization vectors $P$ (Figure 5F). As the size of the domains increases with decreasing temperature, this dielectric mode is characterized with the higher value of $\delta\varepsilon$.

Finally, we conducted a computational analysis on 3CN and 4CN to explore possible explanations for their mesomorphism. The molecular dipole density and the molecular aspect ratio have been proven to play a crucial role in the induction and stabilization of polar phases in nematic LCs as they affect ferroelectric dipolar interaction and shape anisotropy[41] In this context, we calculated total and components values of dipole moment for optimized geometry using a molecular simulation GAUSSIAN 09 software.[62] For our system, we assumed that the principal molecular axis is along the z-axis and two shorter molecular axes are in the orthogonal directions, as shown in **Figure S2** (Supporting Information).

**Table 2** presents the values of three components of the vector and the total value of dipole moment ($\mu$) of compounds 3CN and 4CN. The vector of molecular dipole moment takes the orientation close to the principal molecular axis, which positively contributes to the LC phase stability and explains a broad temperature range of LC phases of 3CN and 4CN. The elongation of the terminal alkyl chain increases the shape anisotropy of rod-shaped molecules, which increases the tendency for apolar N phase formation.[41] On the other hand, the dipolar density, calculated by dividing the total dipole moment by the length of the molecules, is decreased. However, the total dipole moment value remains nearly constant. Thus, stronger dipolar interaction in 3CN than in 4CN provides the enantiotropic nature of the $N_F$ phase for 3CN.

**Table 2.** Components and the total dipole moment values calculated for the optimized geometry of compounds 3CN and 4CN.

| **Compound** | **Dipole moment [Debye]** | | | |
|---|---|---|---|---|
| | $\mu_x$ | $\mu_y$ | $\mu_z$ | $\mu_{total}$ |
| 3CN | 4.6 | -0.1 | 12.9 | 13.8 |
| 4CN | 5.1 | -1.0 | 13.2 | 14.2 |

## 3. Conclusion

In summary, we employed complementary experimental and theoretical methods to confirm the existence of ferro- and antiferroelectric nematic phases in two cyano-terminated and fluorinated phenyl esters synthesized in 2009.[51] POM images of the cell with various types of alignment conditions revealed the characteristic textures corresponding to the $N_F$ and $N_X$ phases for both of the studied compounds. The enantiotropic nature of the $N_F$ and $N_X$ phases for 3CN, and the $N_X$ phase for 4CN, was verified through DSC measurements conducted on individual compounds and one mixture with DIO.[17] In contrast to the ferro- and antiferroelectric nematic LCs developed to date, 3CN and 4CN compounds show much broader temperature ranges of the enantiotropic $N_F$ and $N_X$ phases, respectively. Moreover, we conducted dielectric spectroscopy and SHG measurements on ferro- and antiferroelectric nematic liquid crystals for the first time during the heating cycle. These measurements confirmed the enantiotropic nature of two polar nematic phases in 3CN, and one polar nematic phase in 4CN, as well as their ferro- and antiferroelectric features. Additionally, we present evidence derived from the computational analysis, suggesting that stronger dipolar interactions may be responsible for the enantiotropic nature of $N_F$ in 3CN, as compared to 4CN. This work makes a conceptual advance in designing NLC compounds to achieve the ferro- and antiferroelectric nematic phases with high thermodynamic stability, which can pave the way for room-temperature polar nematics with high application potential in photonic devices.

## 4. Experimental section

*Molecular Modeling.* The dipole moment values for the optimized geometry were calculated using a molecular simulation GAUSSIAN 09 software.[62] Optimization of the molecular structure and other calculations were conducted using the combination of B3LYP (Becke's three parameter hybrid functional using the LYP correlation functional) with the 6-311G plus (d,p) basis set as was detailed described in our previous paper.[54] A frequency check was used to confirm that the minimum energy conformation found was an energetic minimum.

*Differential Scanning Calorimetry.* For differential scanning calorimetry (DSC) studies, a SETARAM DSC 141 calorimeter was used, calibrated using indium and zinc standards. Heating and cooling rates were 2.0 K min$^{-1}$, and samples were kept in a nitrogen atmosphere with a gas flow rate of 20.0 ml min$^{-1}$. The transition temperatures and associated thermal effects were extracted from the heating or cooling traces.

*Dielectric Spectroscopy.* The dielectric measurements were performed using an impedance analyzer, Hewlett Packard 4192A, in a wide frequency range from 100 Hz to 10 MHz. The measuring field (AC) was 0.1 V. Complex electric permittivity of the materials was obtained in 10.0 ± 0.1 μm thick cells with ITO electrodes of low resistivity ($\rho \sim 10$ Ω/sq). The inner surface of the electrodes was covered by antiparallel rubbed polyimide SE-130 (Nissan Chemicals) to provide planar alignment in the cells. The cell temperature was controlled by Linkam TMS 92 and hot-stage Linkam TMSH 600 with an accuracy of 0.1°C.

*Polarized Optical Microscopy.* The optical studies were conducted using a polarized optical microscope Jenapol, Carl Zeiss Jena. The microscope was equipped with a Linkam TMS93 temperature controller with a THMSE 600 heating stage. The materials were sandwiched between untreated glass plates for the observation of "natural" textures or filled at an isotropic phase by capillary action to 8.0 ± 0.1 μm thick cells with planar alignment prepared the same way as for the dielectric measurements.

*Second Harmonic Generation.* For the SHG measurements, a laser beam generated by a Ti:sapphire femtosecond laser amplifier (Spitfire ACE) consisting of 40-femtosecond-long pulses with a central wavelength of 800 nm and a pulse repetition rate of 5 kHz was used. Radiant exposure of each individual pulse was set to ~ 0.01 mJ cm$^{-2}$. The resultant collimated pulse beam reaches the sample placed into a Linkam heating stage equipped with a temperature controller providing an accuracy of ±0.1 K. For SHG measurements, we utilized 7 μm thick planar cells with a polymer layer rubbed in a certain direction to ensure the geometry necessary for the measurements. The cells were positioned with the rubbing direction (long molecular axis) being parallel to the polarization of the incident beam. The SHG signal generated in the sample in a transmission configuration is subsequently spectrally filtered (using optical dichroic mirrors with the central wavelength of 400nm) and finally is detected with an avalanche photodiode and amplified using a lock-in amplifier. The set-up for the temperature-dependent SHG measurements is shown in **Scheme S1** (Supporting Information).

**Supporting Information**

Supporting Information is available from the Wiley Online Library or from the author.

**Data availability**

Data for this article are available in the Zenodo repository at https://doi.org/10.5281/zenodo.14981398.

**Acknowledgments**

MM, MCZ, PP, and DW acknowledge the financial support from the NCN MINIATURA project 2022/06/X/ST5/01316, and the MUT University grant UGB 22-720, and UGB 22-723. NP and DR thank the bilateral project MEYS (project number 8J20PL008) and the project FerroFluid, EIG Concert Japan, 9th call "Design of Materials with Atomic Precision". DR also thanks to the Grant Agency of the Czech Technical University in Prague (Project No. SGS22/182/OHK4/3T/14). RJM acknowledges funding from UKRI via a Future Leaders Fellowship, grant no. MR/W006391/1, and funding from the University of Leeds via a University Academic Fellowship.

**Author contributions**

MM measured dielectric spectra, performed optical microscopy measurements, analyzed the data, conceived the project, and wrote the original draft. MC performed molecular modeling, performed DSC measurements, analyzed the data, directed the project, and wrote the original draft. NP measured the SHG signal, analyzed the data, and wrote the corresponding parts of the draft, DR prepared the SHG set-up, helped with SHG measurements, and wrote the corresponding methodology part of the draft. PP analyzed the dielectric data. RJM synthesized the DIO compound. DW synthesized the 3CN and 4CN compounds, performed optical microscopy measurements, analyzed the data, and wrote the corresponding part of the draft. All authors contributed to scientific discussions.

**ORCID identification numbers**

Mateusz Mrukiewicz: 0000-0002-0212-4520

Michał Czerwiński: 0000-0002-9961-535X

Natalia Podoliak: 0000-0002-3876-5696

Dalibor Repček: 0000-0003-2752-8406

Paweł Perkowski: 0000-0001-7960-2191

Richard. J. Mandle 0000-0001-9816-9661

Dorota Węgłowska: 0000-0002-4887-8459

**References**

[1] T. Y. Kim, S. K. Kim, S. W. Kim, *Nano Converg.* **2018**, *5*, 30.

[2] D. Meier, S. M. Selbach, *Nat. Rev. Mater.* **2022**, *7*, 157.

[3] P. Bednyakov, T. Sluka, A. Tagantsev, D. Damjanovic, N. Setter, *Adv. Mater.* **2016**, *28*, 9498.

[4] L. W. Martin, A. M. Rappe, *Nat. Rev. Mater.* **2017**, *2*, 16087.

[5] G. F. Nataf, M. Guennou, G. Scalia, X. Moya, T. D. Wilkinson, J. P. F. Lagerwall, *Appl. Phys. Lett.* **2020**, *116*, 212901.

[6] T. Mikolajick, S. Slesazeck, H. Mulaosmanovic, M. H. Park, S. Fichtner, P. D. Lomenzo, M. Hoffmann, U. Schroeder, *J. Appl. Phys.* **2021**, *129*, 100901.

[7] Q. Liu, S. Cui, R. Bian, E. Pan, G. Cao, W. Li, F. Liu, *ACS Nano* **2024**, *8*, 1778.

[8] R. B. Meyer, L. Liebert, L. Strzelecki, P. Keller, *J. Phys. (Paris), Lett.* **1975**, *36*, 69.

[9] A. M. Levelut, C. Germain, P. Keller, L. Liebert, J. Billard, *J. Phys.* **1983**, *44*, 623.

[10] J. W. Goodby, E. Chin, *Liq. Cryst.* **1988**, *3*, 1245.

[11] A. D. L. Chandani, E. Gorecka, Y. Ouchi, H. Takezoe, A. Fukuda, *Jpn. J. Appl. Phys.* **1989**, *28*, L1265.

[12] T. Niori, T. Sekine, J. Watanabe, T. Furukawa, H. Takezoe, *J. Mater. Chem.* **1996**, *6*, 1231.

[13] M. Born, *Sitzungsber. Preuss. Akad Wiss.* **1916**, *30*, 614.

[14] R. J. Mandle, S. J. Cowling, J. W. Goodby, *Phys. Chem. Chem. Phys.* **2017**, *19*, 11429.

[15] R. J. Mandle, S. J. Cowling, J. W. Goodby, *Chem. - Eur. J.* **2017**, *23*, 14554.

[16] R. J. Mandle, N. Sebastián, J. Martinez-Perdiguero, A. Mertelj, *Nat. Commun.* **2021**, *12*, 4962.

[17] H. Nishikawa, K. Shiroshita, H. Higuchi, Y. Okumura, Y. Haseba, S. I. Yamamoto, K. Sago, H. Kikuchi, *Adv. Mater.* **2017**, *29*, 1702354.

[18] E. Cruickshank, P. Rybak, M. M. Majewska, S. Ramsay, C. Wang, C. Zhu, R. Walker, J. M. D. Storey, C. T. Imrie, E. Gorecka, D. Pociecha, *ACS Omega* **2023**, *8*, 36562.

[19] X. Chen, V. Martinez, E. Korblova, G. Freychet, M. Zhernenkov, M. A. Glaser, C. Wang, C. Zhu, L. Radzihovsky, J. E. Maclennan, D. M. Walba, N. A. Clark, *PNAS* **2023**, *120*, e2217150120.

[20] R. Saha, P. Nepal, C. Feng, M. S. Hossain, M. Fukuto, R. Li, J. T. Gleeson, S. Sprunt, R. J. Twieg, A. Jákli, *Liq. Cryst.* **2022**, *49*, 1784.

[21] A. Manabe, M. Bremer, M. Kraska, *Liq. Cryst.* **2021**, *48*, 1079.

[22] S. Brown, E. Cruickshank, J. M. D Storey, C. T. Imrie, D. Pociecha, M. Majewska, A. Makal, E. Gorecka, *ChemPhysChem* **2021**, *22*, 2506.

[23] D. Pociecha, R. Walker, E. Cruickshank, J. Szydlowska, P. Rybak, A. Makal, J. Matraszek, J. M. Wolska, J. M. D. Storey, C. T. Imrie, E. Gorecka, *J. Mol. Liq.* **2022**, *361*, 119532.

[24] H. Nishikawa, K. Sano, F. Araoka, *Nat. Commun.* **2022**, *13*, 1142.

[25] M. Mrukiewicz, P. Perkowski, J. Karcz, P. Kula, *Phys. Chem. Chem. Phys.* **2023**, *25*, 13061.

[26] N. Sebastián, M. Čopič, A. Mertelj, *Phys. Rev. E.* **2022**, *106*, 021001.

[27] C. L. Folcia, J. Ortega, R. Vidal, T. Sierra, J. Etxebarria, *Liq. Cryst.* **2022**, *49*, 899.

[28] N. Sebastián, M. Lovšin, B. Berteloot, N. Osterman, A. Petelin, R. J. Mandle, S. Aya, M. Huang, I. Drevenšek-Olenik, K. Neyts, A. Mertelj, *Nat. Commun.* **2023**, *14*, 3029.

[29] E. Zavvou, M. Klasen-Memmer, A. Manabe, M. Bremer, A. Eremin, *Soft Matter* **2022**, *18*, 8804.

[30] J. Li, R. Xia, H. Xu, J. Yang, X. Zhang, J. Kougo, H. Lei, S. Dai, H. Huang, G. Zhang, F. Cen, Y. Jiang, S. Aya, M. Huang, *J. Am. Chem. Soc.* **2021**, *143*, 2023.

[31] B. Basnet, M. Rajabi, H. Wang, P. Kumari, K. Thapa, S. Paul, M. O. Lavrentovich, O. D. Lavrentovich, *Nat. Commun.* **2022**, *13*, 3932.
[32] X. Chen, E. Korblova, D. Dong, X. Wei, R. Shao, L. Radzihovsky, M. A. Glaser, J. E. Maclennan, D. Bedrov, D. M. Walba, N. A. Clark, *PNAS* **2020**, *117*, 14021.
[33] H. Nishikawa, F. Araoka, *Adv. Mater.* **2021**, *33*, 2101305.
[34] X. Chen, E. Korblova, M. A. Glaser, J. E. MacLennan, D. M. Walba, N. A. Clark, *PNAS* **2021**, *118*, e2104092118.
[35] N. Sebastián, R. J. Mandle, A. Petelin, A. Eremin, A. Mertelj, *Liq. Cryst.* **2021**, *48*, 2055.
[36] C. Feng, R. Saha, E. Korblova, D. Walba, S. N. Sprunt, A. Jákli, *Adv. Opt. Mater.* **2021**, *9*, 2101230.
[37] X. Chen, Z. Zhu, M. J. Magrini, E. Korblova, C. S. Park, M. A. Glaser, J. E. Maclennan, D. M. Walba, N. A. Clark, *Liq. Cryst.* **2022**, *49*, 1531.
[38] R. J. Mandle, S. J. Cowling, J. W. Goodby, *Liq. Cryst.* **2021**, *48*, 1780.
[39] J. Li, H. Nishikawa, J. Kougo, J. Zhou, S. Dai, W. Tang, X. Zhao, Y. Hisai, M. Huang, S. Aya, *Sci. Adv.* **2021**, *7*, 5047.
[40] X. Zhao, J. Zhou, J. Li, J. Kougo, Z. Wan, M. Huang, S. Aya, *PNAS* **2021**, *118*, e2111101118.
[41] J. Li, Z. Wang, M. Deng, Y. Zhu, X. Zhang, R. Xia, Y. Song, Y. Hisai, S. Aya, M. Huang, *Giant* **2022**, *11*, 100109.
[42] Y. Song, J. Li, R. Xia, H. Xu, X. Zhang, H. Lei, W. Peng, S. Dai, S. Aya, M. Huang, *Phys. Chem. Chem. Phys.* **2022**, *24*, 11536.
[43] S. Dai, J. Li, J. Kougo, H. Lei, S. Aya, M. Huang, *Macromolecules* **2021**, *54*, 6045.
[44] H. Kikuchi, H. Matsukizono, K. Iwamatsu, S. Endo, S. Anan, Y. Okumura, *Adv. Sci.* **2022**, *9*, 1.
[45] R. J. Mandle, *Crystals (Basel)* **2023**, *13*, 857.
[46] M. Cigl, N. Podoliak, T. Landovský, D. Repček, P. Kužel, V. Novotná, *J. Mol. Liq.* **2023**, *385*, 122360.
[47] G. Stepanafas, E. Cruickshank, S. Brown, M. M. Majewska, D. Pociecha, E. Górecka, J. M. D. Storey, C. T. Imrie, *Mater. Adv.* **2024**, *5*, 525.
[48] H. Nishikawa, K. Sano, S. Kurihara, G. Watanabe, A. Nihonyanagi, B. Dhara, F. Araoka, *Commun. Mater.* **2022**, *3*, 89.
[49] A. Mertelj, L. Cmok, N. Sebastián, R. J. Mandle, R. R. Parker, A. C. Whitwood, J. W. Goodby, M. Čopič, *Phys. Rev. X* **2018**, *8*, 041025.

[50] J. Karcz, N. Rychłowicz, M. Czarnecka, A. Kocot, J. Herman, P. Kula, *Chem. Commun.* **2023**, *59*, 14807.

[51] D. Ziobro, J. Dziaduszek, M. Filipowicz, R. Dąbrowski, J. Czub, S. Urban, *Mol. Cryst. Liq.* **2009**, *502*, 258.

[52] D. Ziobro, P. Kula, J. Dziaduszek, M. Filipowicz, R. Dąbrowski, J. Parka, J. Czub, S. Urban, S. T. Wu, *Opto-electron. Rev.* **2009**, *17*, 16.

[53] H. Xianyu, S.-T. Wu, C.-L. Lin, *Liq. Cryst*. **2009**, *36*, 717.

[54] D. Węgłowska, M. Czerwiński, P. Kula, M. Mrukiewicz, R. Mazur, J. Herman, *Fluid Ph. Equilibria* **2020**, *522*, 112770.

[55] M. Mrukiewicz, P. Perkowski, O. Strzezysz, D. Węgłowska, W. Piecek, *Phys. Rev. E.* **2018**, *97*, 052704.

[56] N. Yadav, Y. P. Panarin, J. K. Vij, W. Jiang, G. H. Mehl, *J. Mol. Liq.* **2023**, *378*, 121570.

[57] S. Kumari, I. M. L. Das, R. Dhar, R. Dabrowski, *J. Mol. Liq.* **2012**, *168*, 54.

[58] P. Perkowski, M. Mrukiewicz, M. Żurowska, R. Dąbrowski, & L. Jaroszewicz, *Liq. Cryst.* **2013**, *40*, 864.

[59] J. Hoffman, W. Kuczyński, J. Małecki, *Mol. Cryst. Liq.* **1978**, *44*, 287.

[60] M. Buivydas, F. Gouda, G. Andersson, S. T. Lagerwall, B. Stebler, J. Bo, È. Melburg, G. Heppke, B. Gestblom, *Liq. Cryst.* **1997**, *23*, 723.

[61] K. S. Cole, R. H. Cole, *J. Chem. Phys.* **1941**, *9*, 341.

[62] M. J. Frisch, G. W. Trucks, H. B. Schlegel, G. E. Scuseria, M. A. Robb, J. R. Cheeseman, G. Scalmani, V. Barone, B. Mennucci, G. A. Petersson, H. Nakatsuji, M. Caricato, X. Li, H. P. Hratchian, A. F. Izmaylov, J. Bloino, G. Zheng, J. L. Sonnenberg, M. Hada, M. Ehara, K. Toyota, R. Fukuda, J. Hasegawa, M. Ishida, T. Nakajima, Y. Honda, O. Kitao, H. Nakai, T. Vreven, J. A. Montgomery, J. E. Peralta, F. Ogliaro, M. Bearpark, J. J. Heyd, E. Brothers, K. N. Kudin, V. N. Staroverov, R. Kobayashi, J. Normand, K. Raghavachari, A. Rendell, J. C. Burant, S. S. Iyengar, J. Tomasi, M. Cossi, N. Rega, J. M. Millam, M. Klene, J. E. Knox, J. B. Cross, V. Bakken, C. Adamo, J. Jaramillo, R. Gomperts, R. E. Stratmann, O. Yazyev, A. J. Austin, R. Cammi, C. Pomelli, J. W. Ochterski, R. L. Martin, K. Morokuma, V. G. Zakrzewski, G. A. Voth, P. Salvador, J. J. Dannenberg, S. Dapprich, A. D. Daniels, J. B. Farkas Foresman, J. V. Ortiz, J. Cioslowski and D. J. Fox, *Gaussian 09, Revision B.01, Gaussian, Inc., Wallingford CT.* **2010**

*Supporting Information*

**Polar Nematic Phases with Enantiotropic Ferro- and Antiferroelectric Behavior**

*Mateusz Mrukiewicz, Michał Czerwiński*, Natalia Podoliak, Dalibor Repček, Paweł Perkowski, Richard. J. Mandle, Dorota Węgłowska*

M. Mrukiewicz, P. Perkowski
Institute of Applied Physics, Military University of Technology, Kaliskiego 2, 00-908 Warsaw, Poland

M.Czerwiński, D. Węgłowska
Institute of Chemistry, Military University of Technology, Kaliskiego 2, 00-908 Warsaw, Poland
E-mail: michal.czerwinski@wat.edu.pl

N. Podoliak, D. Repček
Institute of Physics, Academy of Science of the Czech Republic, Na Slovance 2, 182 00 Prague 8, Czech Republic

D. Repček
Faculty of Nuclear Sciences and Physical Engineering, Czech Technical University in Prague, Břehová 7, 110 00 Prague 1, Czech Republic

R. J. Mandle
School of Chemistry, University of Leeds, Leeds, UK, LS2 9JT
School of Physics and Astronomy, University of Leeds, Leeds, UK, LS2 9JT

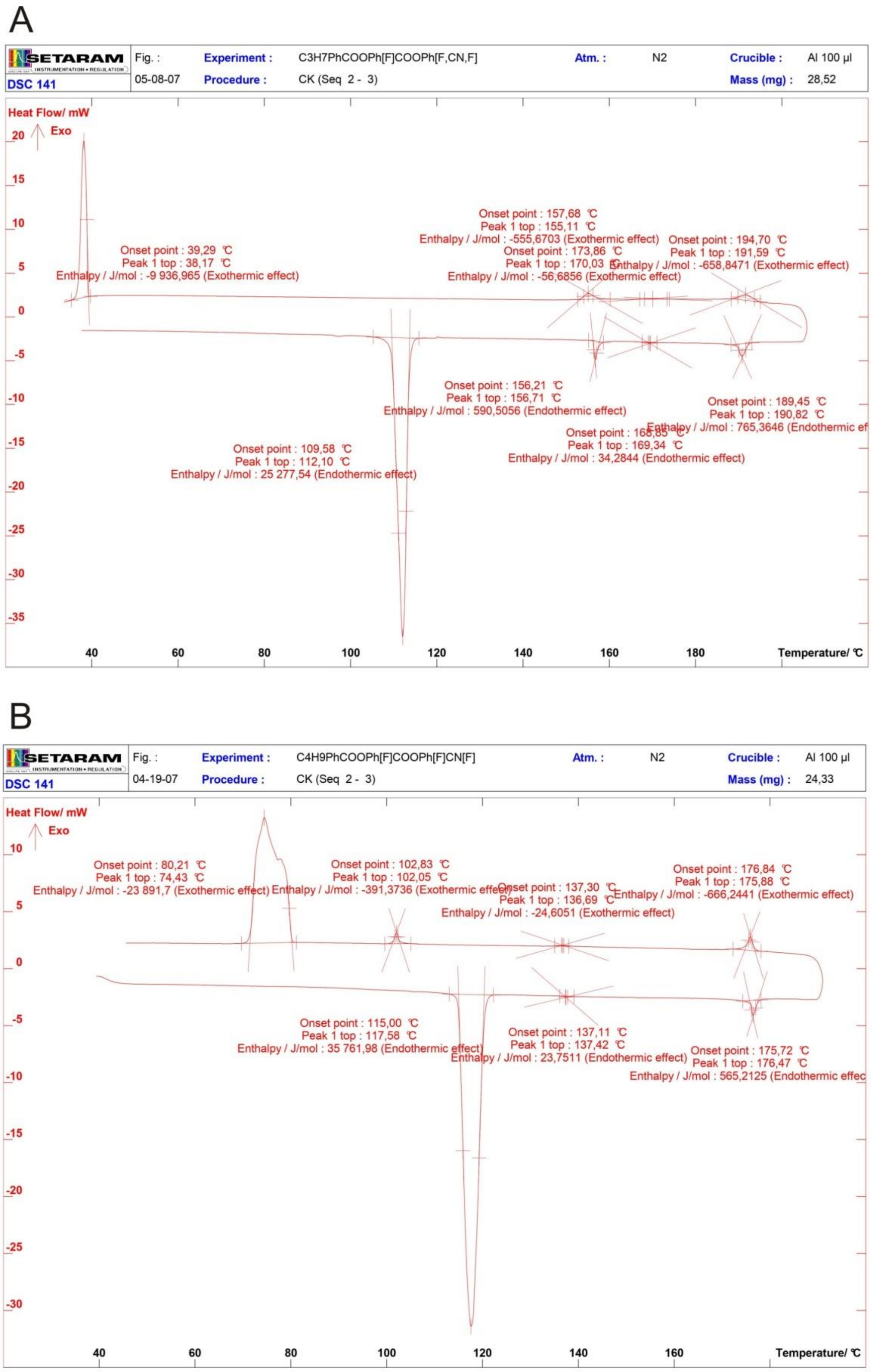

**Figure S1.** The DSC thermograms of compound 3CN (A) and 4CN (B) in the heating cycles (down curves) and cooling cycles (upper curves).

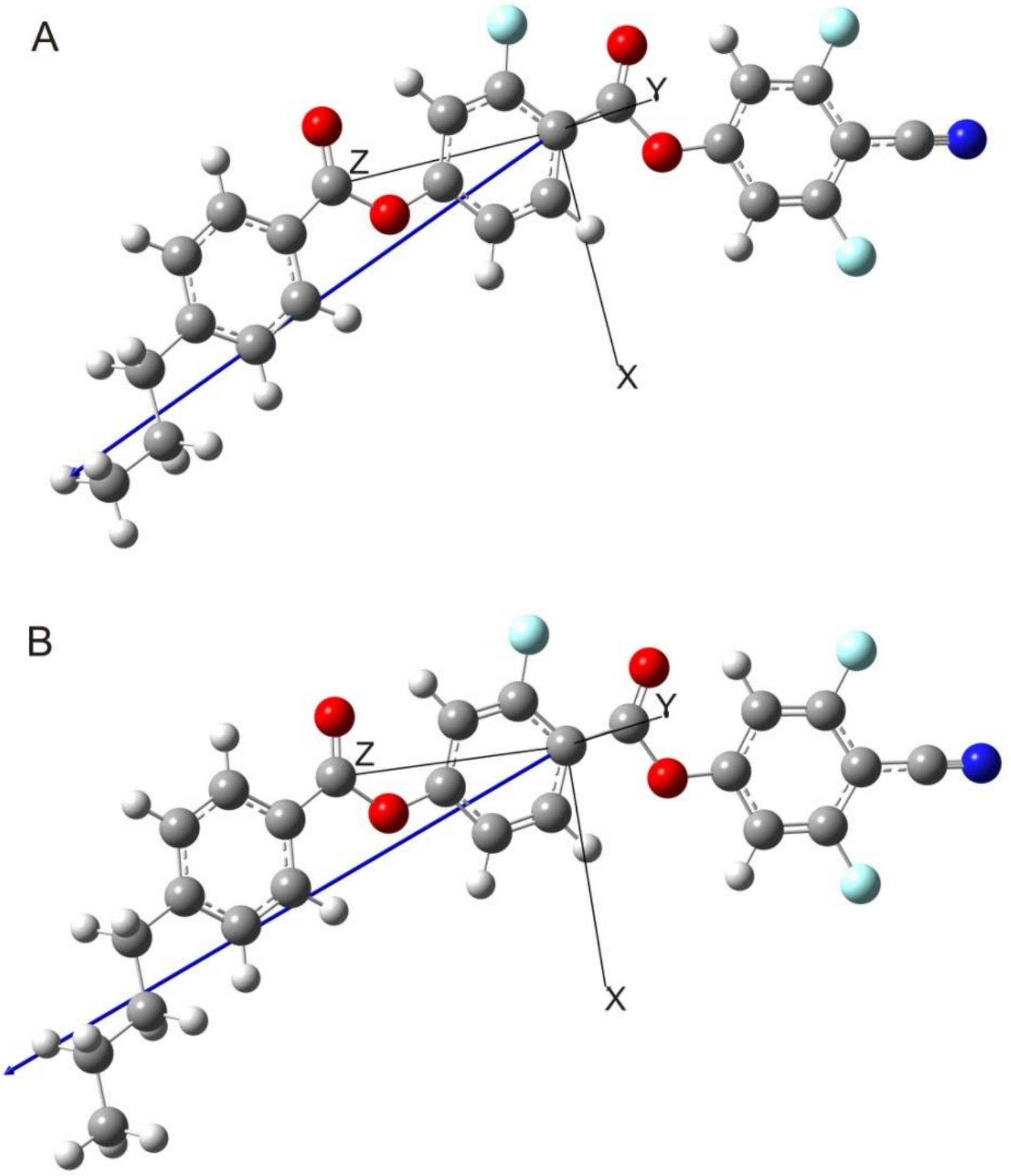

**Figure S2**. The optimized geometric general structure of compounds 3CN (A) and 4CN (B); whereas *x* is the axis in the plane of the phenyl rings, *y* - the axis perpendicular to the plane of the phenyl rings and *z*-the axis along to the principal molecular axis.

VIDEO-S1

**Video S1.** The movie shows the texture change obtained for material between untreated glass plates under cross polarizers during cooling from the nematic (N) to the ferroelectric nematic ($N_F$) phases for 3CN compound.

VIDEO-S2

**Video S2.** The movie shows the texture change obtained for planarly aligned material after applying a triangular in-plane wave electric field of 2 Vpp/20 μm with a frequency of 100 mHz in the $N_F$ phase of the 3CN compound.

VIDEO-S3

**Video S3.** The movie shows the texture change obtained for planarly aligned material after applying a triangular in-plane wave electric field of 1.2 Vpp/20 μm with a frequency of 100 mHz in the $N_F$ phase of the 4CN compound.

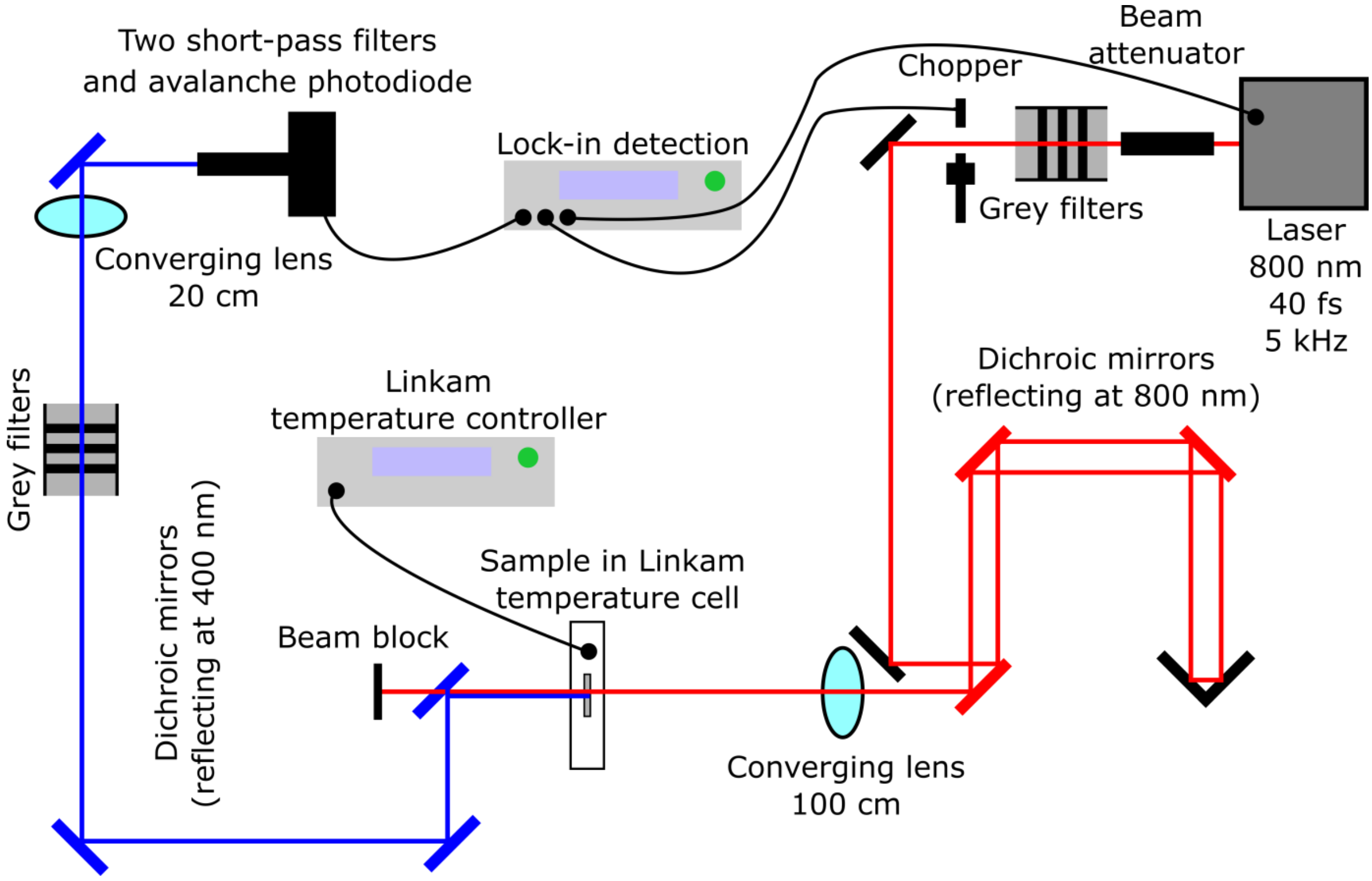

**Scheme S1.** The set-up for the temperature-dependent second harmonic generation (SHG) measurements.